\documentclass[aps,twocolumn,pra,superscriptaddress,amsmath,amssymb]{revtex4-1}
\usepackage{dcolumn}
\usepackage{bm}
\usepackage{amsmath}
\usepackage{txfonts}
\usepackage[T1]{fontenc}
\usepackage{xspace}
\usepackage{ulem}
\usepackage{comment}
\usepackage{braket}
\ifx\pdfoutput\undefined
\usepackage[dvipdfmx]{graphicx}
\usepackage[dvipdfmx]{hyperref}
\usepackage[dvipdfmx]{color}
\else
\usepackage{graphicx}
\usepackage{hyperref}
\usepackage{color}
\fi

\begin{document}
\let\emph\textit

\title{
  Quantum algorithm for the microcanonical Thermal Pure Quantum method
}

\author{Kaito Mizukami}
\author{Akihisa Koga}

\affiliation{
  Department of Physics, Tokyo Institute of Technology,
  Meguro, Tokyo 152-8551, Japan
}

\date{\today}
\begin{abstract}
  We present a quantum algorithm
  for the microcanonical thermal pure quantum (TPQ) method,
  which has an advantage in evaluating thermodynamic quantities
  at finite temperatures,
  by combining with some recently developed techniques derived from quantum singular value transformation.
  When the ground energy of quantum systems has already been obtained precisely, 
  the multiple products of the Hamiltonian are efficiently realized
  and the TPQ states at low temperatures are systematically constructed
  in quantum computations.
\end{abstract}
\maketitle

\section{Introduction}
Quantum computer has been considered as a potential tool
over the classical computations.
One of the advantages in the quantum computations
is the reduction of the computational
cost to solve certain problems
{\it e.g.} the prime factorizations~\cite{Shor}
and linear algebraic calculations~\cite{Harrow,Childs}.
Recently, many of such quantum algorithms have been unified together 
by a novel technique known as the quantum singular value transformation (QSVT)~\cite{GrandUnif,QSVT}, 
which allows one to perform a polynomial transformation of the singular values of a matrix embedded in a unitary matrix.
This technique is expected to further accelerate the development of quantum algorithms.
In the field of the condensed matter physics,
the reduction of the computational memory representing the quantum state
is also important to simulate the quantum many-body systems.
When the $S=1/2$ quantum spin model with the finite system size $N$
is considered, 
each quantum state is represented by the vector with $2^N$ elements,
which makes it hard to deal with the larger system
on the classical computers.
On the other hand, as for the quantum computer,
each state is represented in terms of only $N$ qubits.
Since more than 100 qubits have been reported to be realized~\cite{Arute_2019,Ball_2021},
quantum computations are potential candidates
for simulating the quantum systems in the thermodynamic limit.

One of the important applications is
the simulation for the thermodynamic quantities at finite temperatures.
It is known that, in the classical computations,
the quantum Monte Carlo method is one of the powerful methods
to treat the large system
since thermodynamic quantities are evaluated
by the random samplings in $N$ spins.
However, in the frustrated systems,
serious minus sign problems appear at low temperatures,
and thereby it is still hard to examine thermodynamic properties except for
the special cases~\cite{Nakamura,Nasu}.
Therefore, another tool is desired to discuss thermodynamic properties
in the generic quantum spin systems with large clusters.

The Gibbs sampling algorithm on the quantum computer
has been proposed~\cite{Chowdhury}, where
the Gibbs state can be efficiently prepared.
Another complementary method is
the thermal pure quantum (TPQ) method~\cite{mTPQ,cTPQ}, 
where the thermal averages for the physical quantities
are efficiently evaluated with a typical quantum state for the thermal equilibrium at finite temperature.
Its quantum algorithm has recently been developed,
where the TPQ states are represented by means of
the imaginary time evolution~\cite{Powers,Coopmans,gageTPQ}.
These two algorithms are based on the canonical ensemble in the statistical mechanics.
On the other hand, the TPQ states for isolated systems,
whose properties are described by the microcanonical ensemble,
should be important~\cite{mTPQ_SEKI},
{\it e.g.} the effects of the disorders and real-time dynamics in finite systems.
Therefore, as a complemental method,
it is also instructive to construct the TPQ states in isolated systems
by means of the quantum algorithm.

In the manuscript, we present a quantum algorithm
for the microcanonical TPQ method~\cite{mTPQ}. 
In our scheme, a multiple product of the Hamiltonian 
for constructing the TPQ states is realized, by
making advantages of some recently developed techniques derived from QSVT. 
We demonstrate that the squared norm of the TPQ state,
deeply related to the complexity for the quantum simulations, 
decreases with increasing the number of iterations,
but reaches a certain reasonable value
if the precise value of the ground state energy is given as an input of parameters.
This enables us to explore thermodynamic properties of quantum spin systems with the quantum computer.

The paper is organized as follows.
In Sec.~\ref{sec:mTPQ}, we briefly explain the TPQ method.
In Sec.~\ref{sec:tech}, we explain quantum techniques
used in our scheme.
We introduce our TPQ scheme and clarify its complexity in Sec.~\ref{sec:qTPQ}.
Some numerical results for the frustrated spin systems are also addressed.
A summary is given in the last section.

\section{Thermal pure quantum method}\label{sec:mTPQ}
We consider an isolated quantum spin system with the lattice sites $N$,
which is described by a Hamiltonian $H$, 
and assume that the dimension of the Hilbert space is $D=2^N$. 
The TPQ state $\ket{\psi}$ is one of the typical states for a certain temperature, and
the average in the equilibrium state for an operator $A$ is simply evaluated as
\begin{eqnarray}
  \langle A\rangle=\frac{\bra{\psi}A\ket{\psi}}{\braket{\psi|\psi}}.\label{aveA}
\end{eqnarray}
This formula is exact in the thermodynamic limit
as far as $A$ is represented by low-degree polynomials of the local operators.
It is known that, even in the small clusters,
the TPQ method reasonably describes thermodynamic properties in the thermodynamic limit.
An important point is that, in the TPQ method,
one obtains the average without the diagonalization of the Hamiltonian $H$.
Therefore, it has recently been applied to interesting systems
such as the Heisenberg model on frustrated lattices
~\cite{mTPQ,cTPQ, Yamaji_2016, Endo_2018, Suzuki_2019, Schaefer, Shimokawa}
and the Kitaev models
~\cite{Tomishige_2018, Nakauchi_2018, KogaS1_2018, Oitmaa_2018, KogaMix_2019, Hickey_2019, Morita_2020,Taguchi2022}
to discuss their thermodynamic properties.

Now, we briefly explain the microcanonical TPQ method~\cite{mTPQ}.
Here, we denote the minimum and maximum eigenvalue of the Hamiltonian $H$ 
by $E_\mathrm{min}$ and $E_\mathrm{max}$, respectively.
A TPQ state at $T\rightarrow\infty$ is simply given by a random state
\begin{eqnarray}
  \ket{\psi_0}=\sum_{i=1}^{D} c_i \ket{i},\label{psi0}
\end{eqnarray}
where $\{c_i\}$ is a set of random complex numbers satisfying $\sum_i |c_i|^2 = 1$ and
$\ket{i}$ is an arbitrary orthonormal basis.
By multiplying a certain TPQ state by the Hamiltonian, the TPQ states
at lower temperatures are constructed.
Then, the $k$th TPQ state is represented as \cite{mTPQ}
\begin{eqnarray}
  \ket{\psi_k}= (L-H) \ket{\psi_{k-1}}=(L-H)^k \ket{\psi_{0}},\label{TPQ}
\end{eqnarray}
where $L(>E_\mathrm{max})$ is a constant value.
The internal energy ${\cal E}$ and inverse temperature $\beta$ are given as,
\begin{eqnarray}
  {\cal E}_k&=&\frac{\braket{\psi_k|H|\psi_k}}{\braket{\psi_k|\psi_k}},\\
  \beta_k &=& \frac{2k}{L-{\cal E}_k}.\label{eq:beta}
\end{eqnarray}
Since the temperature has an intensive property,
the number of the iterations $k$, which is proportional to the system size $N$,
is needed to access a certain temperature.
Thermodynamic quantities such as the specific heat and entropy
are evaluated from the above quantities 
and the average of the operator $A$ at the temperature $T_k(=1/\beta_k)$
is obtained in eq.~(\ref{aveA}).
Since the errors in the above formula decrease over the system size, 
a set of the TPQ states generated from a single initial state suffices 
for exploring thermodynamic properties 
in a sufficiently large system
at finite temperatures.

A key of this method is that
the TPQ states are iteratively constructed
in terms of eq.~(\ref{TPQ}).
In the classical computation,
the product between the Hamiltonian and TPQ state is easy to implement.
By contrast, each state is described by the complex vector with $D$ elements
and thereby the feasible cluster size is restricted
by the memory of the classical computer.
On the other hand,
one meets a distinct difficulty in quantum computations.
Each TPQ state can be represented in terms of only $N$ qubits,
while multiple products in eq.~(\ref{TPQ}) exponentially reduce
the success probability in the TPQ method, which will be discussed later.
This means that the simple TPQ simulation is hard to examine
thermodynamic properties at low temperatures.
In the following, combining with some techniques proposed recently,
we present the efficient TPQ scheme to examine thermodynamic properties
on the quantum computer.

\section{main techniques}\label{sec:tech}

In this section,
we explain several techniques based on the QSVT.
In quantum computations,
any operations should be described by the unitary operators.
To operate the Hermite Hamiltonian $H'(=L-H)$,
we use the block-encoding technique.
We here define a unitary matrix $U$,
introducing the $N_a$-qubit ancillary register, as
\begin{eqnarray}
  (\bra{0}_a\otimes I_{s}) U (\ket{0}_a\otimes I_s) = \frac{H'}{\alpha},\label{U}
\end{eqnarray}
where the index $s$ ($a$) represents the system (ancillary) register,
$I$ is an identity matrix, and
$\alpha>\|H'\|$ ($=L-E_\mathrm{min}$) is a positive constant.
Note that $U$, $N_a$ and $\alpha$ 
are not uniquely determined since they depend on the block-encoding technique.
Then, various methods have been proposed~\cite{QSVT,hamilsim}.
One of them is the linear combination of unitaries (LCU) method~\cite{hamilsim}.
This method is applicable for the Hamiltonian,
which is given by $H' = \sum_{j=1}^{N_U} \alpha_j U_j$,
where $U_j$ is a unitary operator, $N_U$ is the number of unitary operators, and 
$\alpha_j$ is a positive constant.
When each $U_j$ is implemented with $C$ primitive gates, 
the unitary $U$ encoding the Hamiltonian requires
$\mathcal{O}(N_UC)$ primitive gates and $\mathcal{O}(\log{N_U})$ ancillary qubits.
In this case, the constant is given as $\alpha=\sum_j \alpha_j$~\cite{hamilsim}.
In addition, another block-encoding method for general sparse matrices has also been proposed~\cite{hamilsim,QSVT}.
In general, the Hermite operator $H'$ can be described
by means of the encoding technique.
For simplicity,
we assume that its gate complexity is given as $C_U$.

In quantum computations with the unitary $U$,
the simple iterative procedure eq.~(\ref{TPQ})
may not be appropriate to construct the $k$th TPQ state.
One of the reasons is the exponential decay
in the amplitude of the TPQ state since $\|H'/\alpha\|<1$.
The other is that the phase shifts of the ancillary qubits
for each unitary operation eq.~(\ref{U})
yields unphysical results in the system registers
after the multiple iterations.
To overcome two problems,
we make use of the uniform spectral amplification and
quantum eigenvalue transformation (QET) techniques.

First, we use the uniform spectral amplification method
to avoid the exponential decay in the amplitude of the TPQ state.
According to the Theorem 30 in Ref.~\cite{QSVT} ( which is a generalization of the Theorem 2 in Ref.~\cite{speamp}) ,
for any $\Lambda \in (\|H'\|,\alpha]$,
$\delta\in \left(0,1-\frac{\|H'\|}{\Lambda}\right]$, and $\epsilon\in (0,1/2)$,
there exists a unitary $\Tilde{U}$ such that
\begin{eqnarray}
  (\bra{0}_a\otimes I_{s}) \Tilde{U} (\ket{0}_a\otimes I_s)
  = \frac{\Tilde{H}'}{\Lambda},
\end{eqnarray}
where $\tilde{H}'$ is 
the approximate matrix of $H'$.
When the error between the $i$th eigenvalues $\Tilde{E}'_i$ and $E'_i$
for $\tilde{H}'$ and $H'$ is bounded as
\begin{eqnarray}
  \frac{|\Tilde{E}'_i-E'_i|}{\Lambda}<\epsilon, \label{eq:error}
\end{eqnarray}
the gate complexity of $\Tilde{U}$ is given as 
\begin{eqnarray}
  \mathcal{O}\left(\frac{\alpha}{\Lambda}\frac{1}{\delta}\log\Biggl(\frac{\alpha}{\Lambda}\frac{1}{\epsilon}\Biggr)\cdot C_U\right).
\end{eqnarray}
We also make use of the QET~\cite{GrandUnif,hamilsim,QSVT}
to perform the multiple operations $\tilde{H}'^k$ exactly.
Let $V$ be a unitary satisfying
\begin{eqnarray}
  (\bra{0}_a\otimes I_{s}) V (\ket{0}_a\otimes I_s)
  = \left(\frac{\Tilde{H}'}{\Lambda}\right)^k,
\end{eqnarray}
and this operation requires two additional ancillary qubits,
$k$ controlled-$\tilde{U}$ gates,
and $\mathcal{O}(k\log{(N_a)})$ primitive gates~\cite{speamp}.
In this connection, one can use a unitary $\tilde{U}$
instead of controlled-$\tilde{U}$ gates when $k$ is odd.

If $|0\rangle_a$ is observed in the ancillary $N_a$ qubits,
one obtains the (normalized) approximate $k$th TPQ state
$|\tilde{\psi}_k\rangle/\sqrt{\braket{\tilde{\psi}_k|\tilde{\psi}_k}}$,
where
\begin{eqnarray}
  |\tilde{\psi}_k\rangle&=&\left(\frac{\tilde{H}'}{\Lambda}\right)^k|\psi_0\rangle.
\end{eqnarray}
The success probability, {\it i.e.} the probability of observing $\ket{0}_a$, 
is given as
\begin{eqnarray}
  \langle \tilde{\psi}_k|\tilde{\psi}_k\rangle
  =\sum_n |c_n|^2 \left(\frac{\tilde{E'}_n}{\Lambda}\right)^{2k},\label{eq:average}
\end{eqnarray}
where we have taken the eigenstates $\{\ket{n}\}$ of the Hamiltonian $H$ 
as the basis
of the initial TPQ state $\ket{\psi_0}$.
To discuss thermodynamic properties at low temperatures, 
we roughly evaluate the quantity for sufficiently large $k$ as
\begin{eqnarray}
  \mathcal{O}\left(|c_{\mathrm{min}}|^2\left(\frac{L-E_\mathrm{min}}{\Lambda}\right)^{2k}\right),\label{pk}
\end{eqnarray} 
where we have assumed $\epsilon \ll (L-E_\mathrm{min})/2k\Lambda$ 
to neglect the effect of the approximation.
$c_\mathrm{min}$ is the coefficient of the eigenstate for $E_\mathrm{min}$ 
in the initial random state 
and $|c_\mathrm{min}|^2\sim \mathcal{O}(1/D)$.
Since the squared norm eq.~(\ref{eq:average}) is tiny in any case, we also need to use amplitude amplification technique~\cite{grover,QSVT}
to complete the implementation of $\tilde{H}'^k$.
It is known that
the success probability can be amplified
to a constant although the quantum complexity increases inversely proportional to the square root of its success probability.
Thus, the number of amplitude amplification steps requires as
\begin{eqnarray}
\mathcal{O}\left(\sqrt{D} \left(\frac{\Lambda}{L-E_\mathrm{min}}\right)^k \right), 
\end{eqnarray}
where the gate complexity of each step is the gate complexity of $V$.
Combining with the above techniques hierarchically,
we construct a quantum algorithm for microcanonical TPQ method, which is explicitly shown in the following.

\section{Quantum algorithm}\label{sec:qTPQ}
Our algorithm is efficient to multiply a random state by $\tilde{H}'^k$ 
and obtain the approximate normalized TPQ state on the quantum computer with constant probability.
Specifically, given a precision parameter $\epsilon'\in (0,\mathcal{O}(L-E_\mathrm{max})]$, one can construct a unitary $\Bar{V}$ that satisfies
\begin{align}
    \frac{(\bra{0}_a\otimes I_s)\ \Bar{V}\ (\ket{0}_a\otimes \ket{\psi_0})}{\|(\bra{0}_a\otimes I_s)\ \Bar{V}\ (\ket{0}_a\otimes \ket{\psi_0})\|}=\frac{\ket{\tilde{\psi}_k}}{\sqrt{\braket{\tilde{\psi}_k|\tilde{\psi}_k}}}
\end{align}
and
\begin{align}
  \left\| \frac{|\tilde{\psi_k}\rangle\langle\tilde{\psi_k}|}
          {\langle\tilde{\psi_k}|\tilde{\psi}_k\rangle}
          - \frac{|\psi_k\rangle\langle\psi_k|}
          {\langle\psi_k|\psi_k\rangle}\right\|_1<\epsilon'\label{eq:errorH},
\end{align}
by using the uniform spectral amplification, QET, and amplitude amplification techniques. 
The condition on the trace distance in eq.~(\ref{eq:errorH} )
means that no measurement can distinguish between 
$|\tilde{\psi_k}\rangle/\sqrt{\langle\tilde{\psi_k}|\tilde{\psi}_k\rangle}$ and $|\psi_k\rangle/\sqrt{\langle\psi_k|\psi_k\rangle}$
with probability greater than $\epsilon'$~\cite{quant_det}.
From the relation between $\epsilon$ and $\epsilon'$ discussed in Appendix~\ref{A}, 
one can get the TPQ state with the desired precision in eq.~(\ref{eq:errorH}) if the error $\epsilon$ satisfies
\begin{eqnarray}
  \epsilon<\frac{\epsilon'}{4k\Lambda}\frac{\sqrt{\braket{\psi_0|H'^{2k}|\psi_0}}}{\| H'^{k-1}\|}.
\end{eqnarray}
Therefore, we can conclude our algorithm in terms of the gate complexity as follows.

\vspace{2mm}
\noindent
{\bf Theorem}
{\it There exists a quantum circuit $\Bar{V}$ to prepare the $k$th TPQ state approximately,}
{\it and
the gate complexity of the unitary $\Bar{V}$ is then given as  
\begin{align}
  \mathcal{O}\left( \left(\frac{\Lambda}{L-E_\mathrm{min}}\right)^k
  \sqrt{D}\cdot k\cdot \frac{\alpha}{\Lambda} \frac{1}{\delta}\log \Bigg(\frac{k\alpha}{\epsilon'}\frac{\| H'^{k-1}\|_1}{\sqrt{\braket{\psi_0|H'^{2k}|\psi_0}}}\Bigg)\cdot C_U\right).
\end{align}
}
\vspace{2mm}
Replacing $\braket{\psi_0|H'^{2k}|\psi_0}$ by its average, we get
\begin{align}
  \frac{\| H'^{k-1}\|_1}{\sqrt{\braket{\psi_0|H'^{2k}|\psi_0}}}\sim \frac{\sum_n {E'_n}^{k-1}}{\sqrt{\frac{1}{D}\sum_n {E'_n}^{2k}}}< \frac{D}{l-E_\mathrm{max}}.
\end{align}
Thus, the gate complexity follows as
\begin{align}
  \mathcal{O}\left( \left(\frac{\Lambda}{L-E_\mathrm{min}}\right)^k
  \sqrt{D}\cdot k\cdot \frac{\alpha}{\Lambda} \frac{1}{\delta}\log \Bigg(\frac{k\alpha}{\epsilon'}\frac{D}{l-E_\mathrm{max}}\Bigg)\cdot C_U\right).
\end{align}
In addition, if the ground state energy has already been obtained precisely, 
we can set $\Lambda=L-E_\mathrm{min}(1\pm \delta')$ where $\delta'>0$ is a precision parameter, and the sign is taken 
so that $\Lambda > L-E_\mathrm{min}$.
When
$\delta'$ is chosen to be small $[\delta' \ll 1/k]$,
the exponential increase with respect to $k$ can be neglected and 
the gate complexity is then given as 
\begin{eqnarray}
  \mathcal{O}\left(
  \sqrt{D} k \frac{\alpha}{\Lambda} \frac{1}{\delta}\log \Bigg(\frac{k\alpha}{\epsilon'}\frac{D}{L-E_\mathrm{max}}\Bigg)C_U\right).
\end{eqnarray}
This means that the precision of the ground energy of the Hamiltonian plays a crucial role
and the exponential increase of the complexity can be suppressed in the case with $\delta' = \mathcal{O}(1/k)$.

More specifically, in the case of the $S=1/2$ quantum spin systems, where the Hamiltonian is, in general,
represented by linear combination of tensor products of Pauli operators,
we have $\alpha=\mathcal{O}(N)$, $N_U = \mathcal{O}(N)$, and $C = \mathcal{O}(1)$ by means of the LCU method.
Note that $\delta=\mathcal{O}(1/k)$, $\Lambda=\mathcal{O}(N)$ and $k=\mathcal{O}(N)$, 
the gate complexity is given as 
\begin{eqnarray}
  \mathcal{O}\left(\sqrt{D}\;\mathrm{polylog}(D)\ \log\Bigl(\frac{1}{\epsilon'(L-E_\mathrm{max})}\Bigr)\right).
\end{eqnarray}
Since the computational complexity for one iteration of the TPQ method on the classical computer
which is constructed by vector and (sparse) matrix operation is $O(D)$,
a quadratic speedup should be realized more or less except for the polylog factors.
As for the memory, the TPQ state is stored in $N$ qubits,
which is much smaller than the complex vector with $D$ elements
in the classical computation.
Therefore, our scheme has a potential method to evaluate statistical-mechanical quantities 
in large systems.

Our quantum algorithm to obtain the $k$th TPQ state is explicitly shown as follows.
We first set constants $L(> E_\mathrm{max})$ and $\Lambda(\sim L-E_\mathrm{min})$. 
We also set the accuracy parameter $\epsilon= \mathcal{O}(L-E_\mathrm{max})$.
The TPQ method on quantum computers is composed of three basic steps.
The first step is the preparation of the initial random state
$|\psi_0\rangle$, which can be obtained from the application of
a random circuit~\cite{GoogleSupremacy,HydroRandom,GoogleRandom,PseudoRandom}.
The second step is to apply the unitary $\Bar{V}$ to the initial state.
The last step is to measure the ancillary register
to obtain the $k$th TPQ state.
When the state $|0\rangle_a$ is observed in the ancillary qubits,
one obtains the TPQ state $|\tilde{\psi}_k\rangle$
in the system register.

\begin{figure}[htb]
  \centering
  \includegraphics[width=\linewidth]{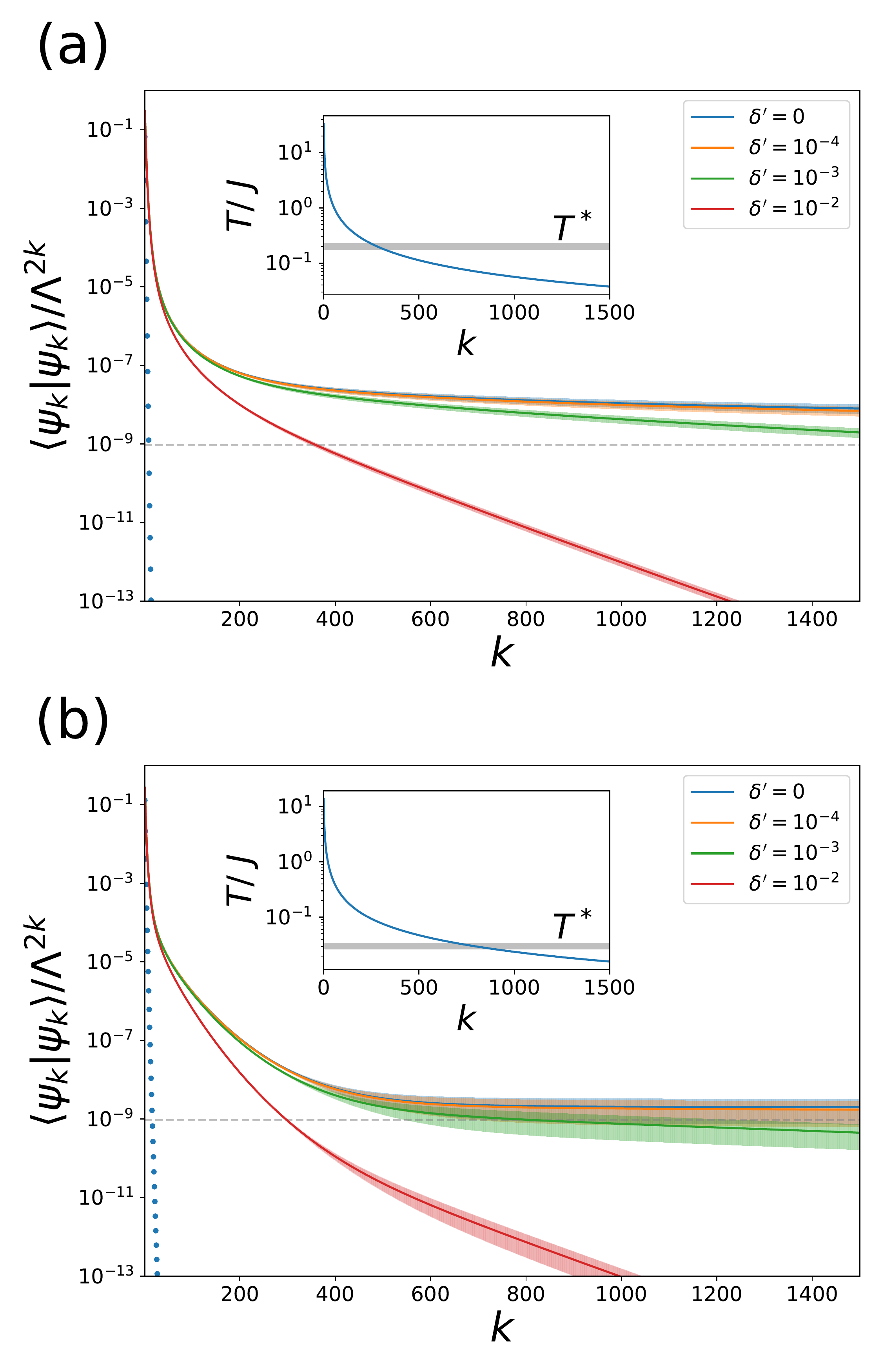}
  \caption{
    The squared norm $\langle {\psi}_k|{\psi}_k \rangle/\Lambda^{2k}$ as a function of $k$
    in (a) the KH and (b) Kitaev models with $N=30$
    when $\delta'=0, 10^{-4}, 10^{-3}$, and $10^{-2}$.
    The shaded areas stand for the standard deviation of the results.
    The TPQ results without the uniform spectral amplification
    is represented by dotted line. 
    The gray dashed line represents $1/D=2^{-30}$.
  }
  \label{suc}
\end{figure}

Now, we calculate the squared norm $\langle {\psi}_k|{\psi}_k \rangle/\Lambda^{2k}$, which is directly related to the complexity of our scheme, 
to demonstrate the quantity is $\mathcal{O}(1/D)$ if $\delta'$ is chosen to be $\mathcal{O}(1/k)$.
To this end, we consider the Heisenberg model on the Kagome lattice (KH model) and Kitaev models with a coupling constant $J$,
as examples of the frustrated quantum spin systems.
In these cases, the corresponding Hamiltonians can be easily encoded by the LCU method with
$\alpha=L+3JN/2$ and $\alpha=L+6JN$, respectively.
The details of these models will be explained in Appendix~\ref{AppendixB}.
Here, the TPQ simulations are performed from 25 independent samples of $|\psi_0\rangle$
on the classical computer, by setting the parameter $L=E_\mathrm{max}+\eta$ with $\eta=0.001N$.
Figure~\ref{suc} shows the squared norm
$\langle {\psi}_k|{\psi}_k \rangle/\Lambda^{2k}$ for both models with $N=30$.
When the bare TPQ method is applied without the uniform spectral amplification technique,
the squared norm rapidly decreases since $L-E_\mathrm{min}<\alpha$.
This is clearly shown as the dotted line in Fig.~\ref{suc}.
On the other hand,
we find that the squared norm decreases slowly with small $\delta'$, 
and is $\mathcal{O}(1/D)$ in a region $k\in [0,\mathcal{O}(1/\delta')]$ for any $\delta'$.
This suggests that preknowledge of $E_\mathrm{min}$ with a precision of $\mathcal{O}(1/k)$
suppresses the exponential decay in the success probability.
The inset of Fig.~\ref{suc} shows the temperature as a function of $k$ in the TPQ simulations.
It is found that the increase of $k$ monotonically decreases the temperature.
In general, there exists the characteristic temperature $T^*$ which depends on the model.
Namely, $T^*\sim 0.2J$ for the KH model and $T^*\sim 0.03J$ for the Kitaev model
(see Appendix~\ref{AppendixB}).
It is found that $k\sim 1500\ (50N)$ iterations are enough to reach the characteristic temperature.
These results imply that the thermodynamic properties can be discussed within a reasonable computational cost. 
It is expected that our quantum scheme is applied to frustrated quantum spin systems and their interesting low-temperature properties are clarified.

\section{Summary}\label{sec:summary}
We have presented the quantum algorithm
for the TPQ method~\cite{mTPQ},
combining with the block-encoding, uniform spectral amplification,
QET, and amplitude amplification techniques.
When the precise value of the ground state energy is given as an input parameter,
the complexity in the multiple products of the Hamiltonian constructing the TPQ states
is exponentially reduced.
This enables us a quadratic speedup except for the polylog factors
compared with the classical simulation.
Furthermore, our quantum scheme has an advantage 
in storing the TPQ state in the computational memory.
Therefore, our work should 
stimulate the further theoretical studies in the condensed matter physics with quantum computer.

\begin{acknowledgments}
We would like to thank K. Fujii for valuable discussions.
This work was supported by Grant-in-Aid for Scientific Research from
JSPS, KAKENHI Grant Nos.
JP22K03525, JP21H01025, JP19H05821 (A.K.).

\end{acknowledgments}

\appendix

\section{Relation between $\epsilon$ and $\epsilon'$} \label{A}

In this section, we consider the relationship between
the approximation error $\epsilon'$ given in eq.~(\ref{eq:error}) and
the error $\epsilon$ between $H'$ and $\Tilde{H}'$ in eq.~(\ref{eq:errorH}),
and evaluate the necessary condition for $\epsilon$ to achieve the desired precision $\epsilon'$.

First, 
for any state $\ket{\psi}$ which satisfies $\braket{\psi|\psi}=1$,
we define $W_k$ as
\begin{align}
  W_k = \frac{\tilde{H'}^k\ket{\psi}\bra{\psi}\tilde{H'}^k}
          {\tilde{N}_k}
          - \frac{{H'}^k\ket{\psi}\bra{\psi}{H'}^k}{N_k},
\end{align}
where $\tilde{N}_k = \braket{\psi|\tilde{H'}^{2k} |{\psi}}$, and $N_k =\braket{\psi|{H'}^{2k} |{\psi}}$.
Then, we obtain
\begin{align}
  &\quad \frac{1}{2}\| W_k\|_1\\
  &\le \frac{1}{2}\left\{\left\|\frac{1}{\tilde{N}_k}(\tilde{H'}^k-H'^k)\ket{\psi}\bra{\psi}\tilde{H'}^k\right\|_1 + 
    \left\| \frac{1}{N_k}{H'}^k\ket{\psi}\bra{\psi}(H'^k-\tilde{H'}^k)\right\|_1\right\}\\
  &= \frac{1}{2}\left( \frac{1}{\sqrt{\tilde{N}_k}}+ \frac{1}{\sqrt{N_k}}\right) \sum_n \left| {\tilde{E'}_n}^k - {{E'}_n}^k\right|.
\end{align}
From $|\Tilde{E'}_n- E'_n|/\Lambda<\epsilon$,
assuming that $\epsilon\le \frac{4}{9}\frac{l-E_\mathrm{max}}{k\Lambda}$, the following inequations hold,
\begin{align}
  \sum_n \left| {\tilde{E'}_n}^k - {{E'}_n}^k\right| &\le 2k\Lambda\epsilon \left\|H'^{k-1}\right\|_1,\\
  \frac{1}{\sqrt{\tilde{N}_k}} &\le \frac{3}{\sqrt{N_k}}.
\end{align}
Thus, we can obtain the upper bound of the above quantity as
\begin{align}
  \frac{1}{2}\| W_k\|_1 \le 4k\Lambda\epsilon \frac{\|H'^{k-1}\|_1}{\sqrt{N_k}}
\end{align}
Therefore, when $\frac{1}{2}\| W_k\|_1<\epsilon'$, the error $\epsilon$ between the eigenvalues in $H'$ and $\tilde{H}'$ should satisfy
\begin{align}
  \epsilon < \frac{\epsilon'}{4k\Lambda}\frac{\sqrt{N_k}}{\|H'^{k-1}\|_1}.
\end{align}
Here, note that since $\epsilon\le \frac{4}{9}\frac{l-E_\mathrm{max}}{k\Lambda}$ must be satisfied, 
we should choose $\epsilon'$ satisfying $\epsilon' \in (0, \mathcal{O}(L-E_\mathrm{max})]$.

\section{Details of the frustrated quantum spin models }\label{AppendixB}
Here, we explain the details of the models used in the TPQ simulations.
In the frustrated spin systems, low energy states should play an important role
and the characteristic temperatures is relatively low, compared to unfrustrated systems.
In fact, low-temperature peak or shoulder in the specific heat has been discussed
in some systems.
Now, we treat the KH and Kitaev model as examples of the frustrated models.
To make our discussions clear, we set $L=E_\mathrm{max}+\eta$ with $\eta=0.001N$

\subsection{The Heisenberg model on the Kagome lattice}
\begin{figure}[htb]
  \centering
  \includegraphics[scale=0.4]{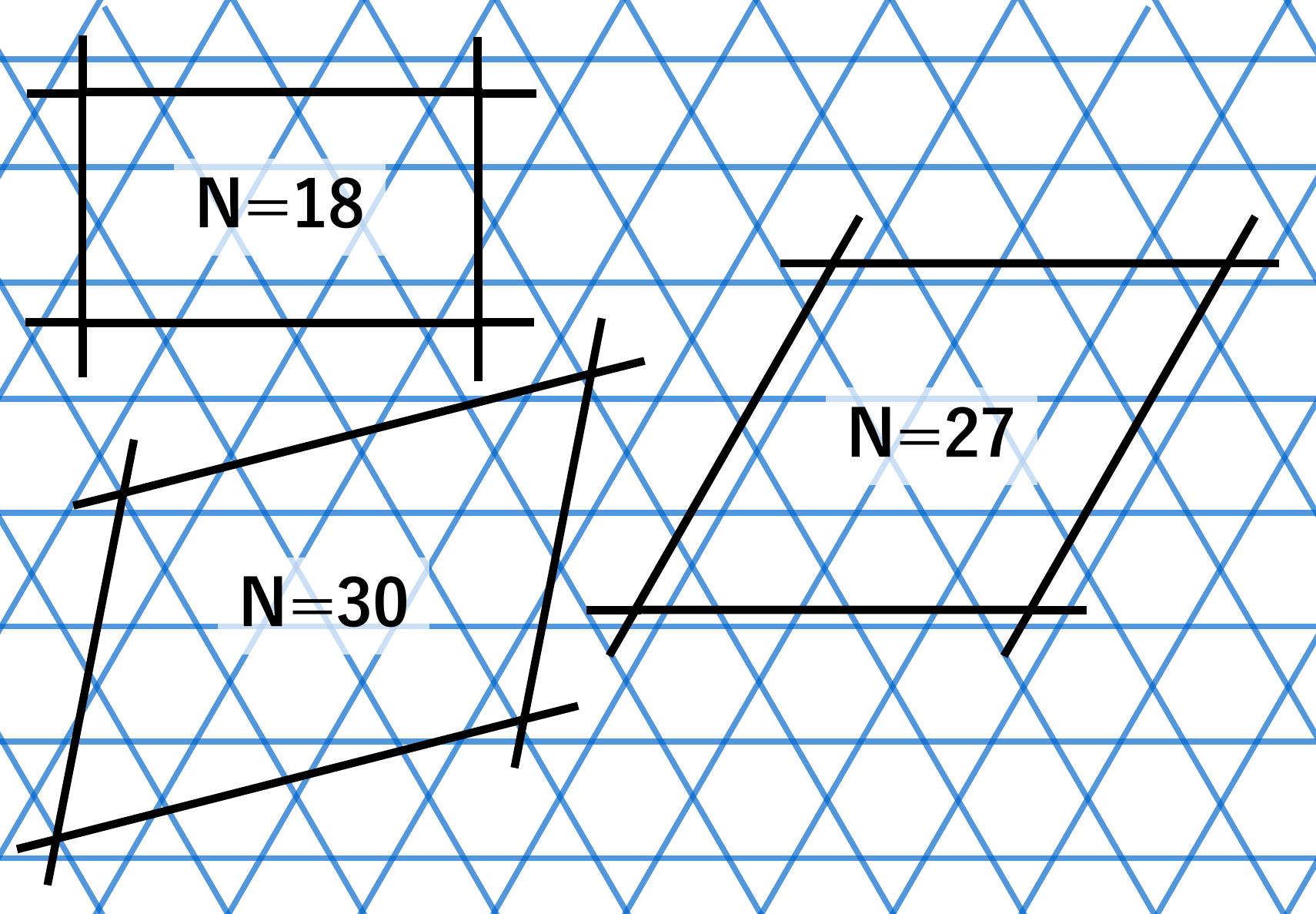}
  \caption{
    Finite size clusters of the Kagome lattice used in the TPQ simulations.
    The boundaries exhibit the periodic boundary conditions.
  }
  \label{fig:Kagome_lattice}
\end{figure}

First, we consider the KH model with antiferromagnetic couplings
as one of the systems with geometrical frustration,
which is schematically shown in Fig.~\ref{fig:Kagome_lattice}.
The system includes triangle structures and each site
connects four nearest neighbor sites.
The model Hamiltonian is given as
\begin{eqnarray}
  H=J\sum_{\langle ij\rangle}{\bf \sigma}_i\cdot{\bf \sigma}_j,
\end{eqnarray}
where ${\bf \sigma}_i[=(\sigma_i^x,\sigma_i^y,\sigma_i^z)]$,
$\sigma_i^\mu$ is the $\mu$ component of the Pauli matrix at the $i$th site and
the index $\langle ij\rangle$ represents the summation over
the connecting spin pairs.
$J(>0)$ is the antiferromagnetic exchange coupling.

\begin{figure}[htb]
  \centering
  \includegraphics[width=\linewidth]{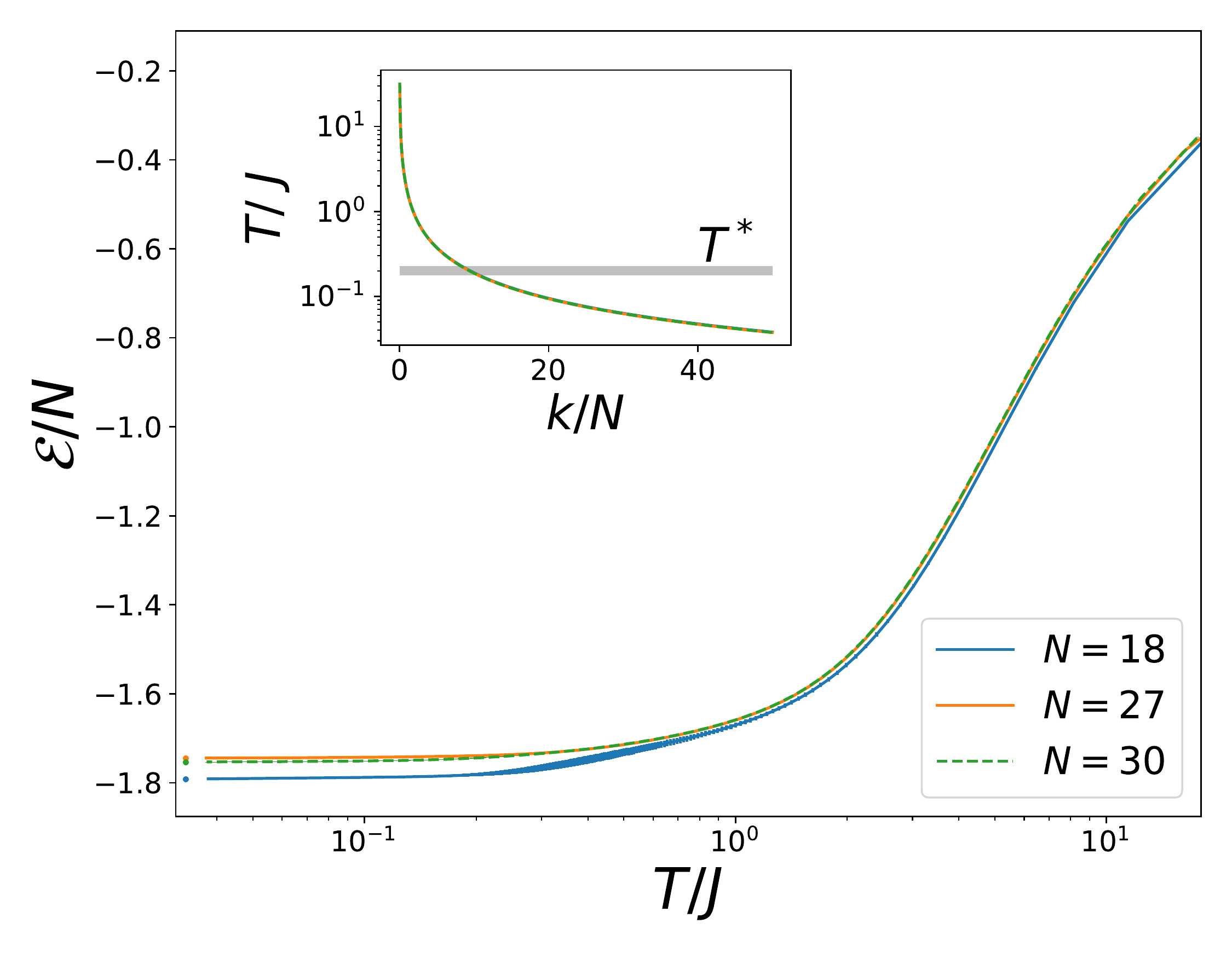}
  \caption{
    Internal energy in the Heisenberg model on the Kagome lattice with $N=18, 27$, and $30$.
    The inset shows the temperature as a function of $k$.
    The error bars stand for the standard deviation of the results.
    Circles represent the ground state energies for the corresponding system.
  }
  \label{fig:Kagome_T}
\end{figure}

For the clusters with $N=18$ and $27$, 
we evaluate temperatures and internal energies by means of $100$ independent TPQ states.
By contrast, the numerical cost is high for the cluster with $N=30$,
and $25$ independent states are treated.
The internal energy is shown in Fig.~\ref{fig:Kagome_T}.
At low temperatures,
the internal energy strongly depends on the size and/or shape of the system.
This means that low energy states play an important role in the Kagome-Heisenberg model.
It has been clarified that there exists shoulder behavior in the specific heat
and its characteristic temperature is deduced as $T^*\sim 0.3J$~\cite{cTPQ}.
The inset of Fig.~\ref{fig:Kagome_T} shows the temperature as a function of the scaled iteration $k/N$.
We find that the curves little depend on $k/N$. 
Therefore, 
the TPQ state at $T=T^*$ is obtained with $k\sim 10N $
when the parameters are appropriately given.

\subsection{The Kitaev model on the honeycomb lattice}
\begin{figure}[htb]
  \centering
  \includegraphics[width=\linewidth]{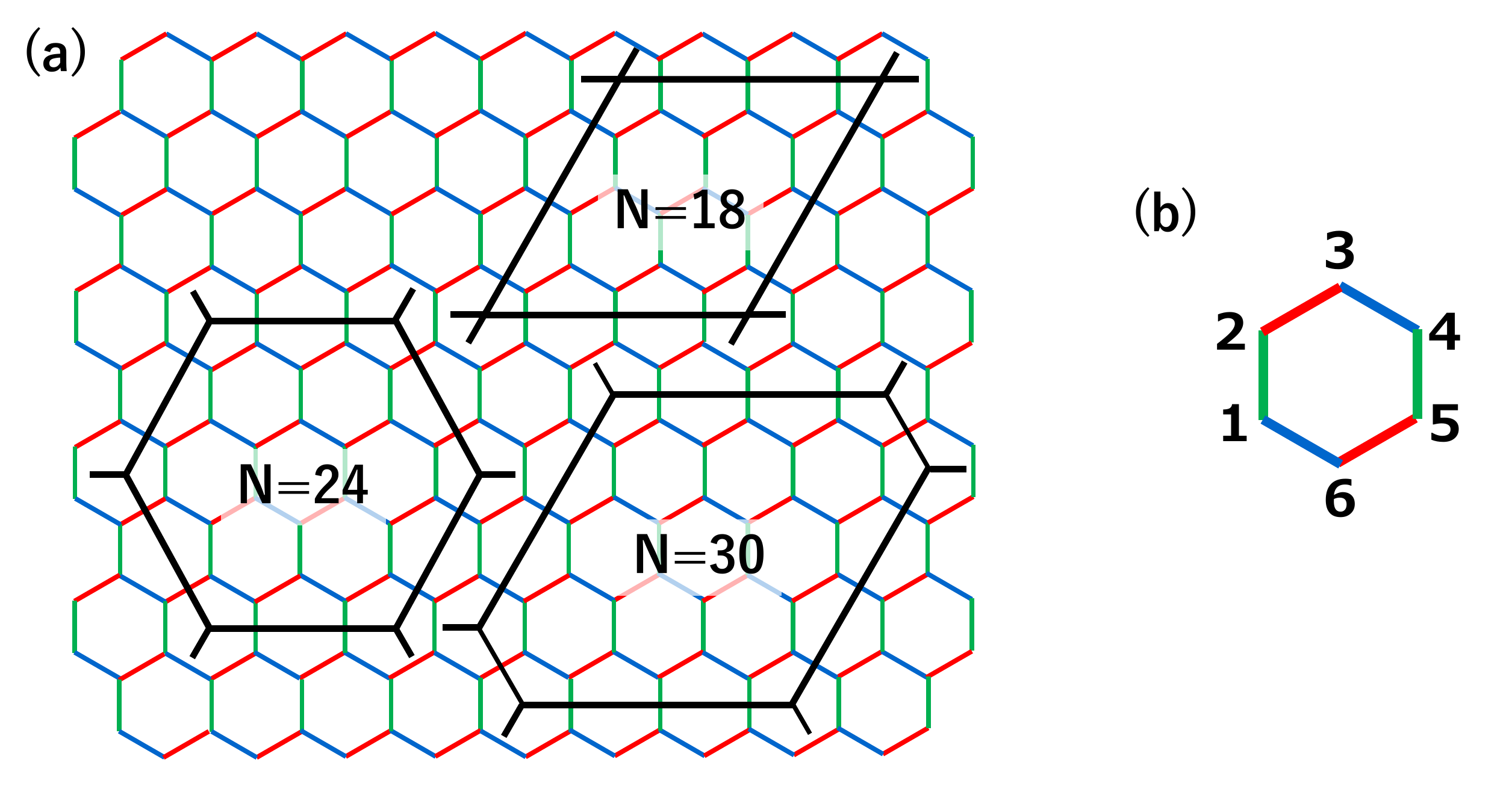}
  \caption{
    (a) Finite size clusters of the Kitaev model on the honeycomb lattice used in the TPQ simulations.
    The boundaries exhibit the periodic boundary conditions.
    Red, blue, and green lines represent $x$-, $y$-, and $z$-bonds, respectively.
    (b) Plaquette with sites marked $1-6$ is shown for
    the corresponding operator $W_p$ (see text).
  }
  \label{fig:lattice}
\end{figure}

We consider the Kitaev model on the honeycomb lattice~\cite{Kitaev},
which is composed of the direction dependent Ising-like interactions and
is known as the exactly solvable systems with bond frustration.
The Hamiltonian is given by
\begin{eqnarray}
  H = -J\sum_{\langle i,j\rangle_x}\sigma_i^x \sigma_j^x
  -J\sum_{\langle i,j\rangle_y}\sigma_i^y \sigma_j^y
  -J\sum_{\langle i,j\rangle_z}\sigma_i^z \sigma_j^z,
\end{eqnarray}
where $\langle i, j\rangle_\mu$ represents
the nearest-neighbor pair on the $\mu(=x,y,z)$-bonds.
The $x$-, $y$-, and $z$-bonds are shown as red, blue, and green lines
in Fig.~\ref{fig:lattice}(a).
$J$ is the exchange coupling between the nearest-neighbor spins.
In the Kitaev model, there exists a local conserved quantity
defined at each plaquette $p$ composed of the sites
labled as $1, 2, \cdots, 6$ [see Fig.~\ref{fig:lattice}(b)]
$W_p=\sigma_{1}^x \sigma_{2}^y \sigma_{3}^z \sigma_{4}^x \sigma_{5}^y \sigma_{6}^z$.
It is known that due to the existence of the local conserved quantities,
the ground state is the quantum spin liquid,
where the spin degrees of freedom is fractionalized into
itinerant Majorana fermions and fluxes.
This leads to two distinct characteristic energy scales.
\begin{figure}[htb]
  \centering
  \includegraphics[width=\linewidth]{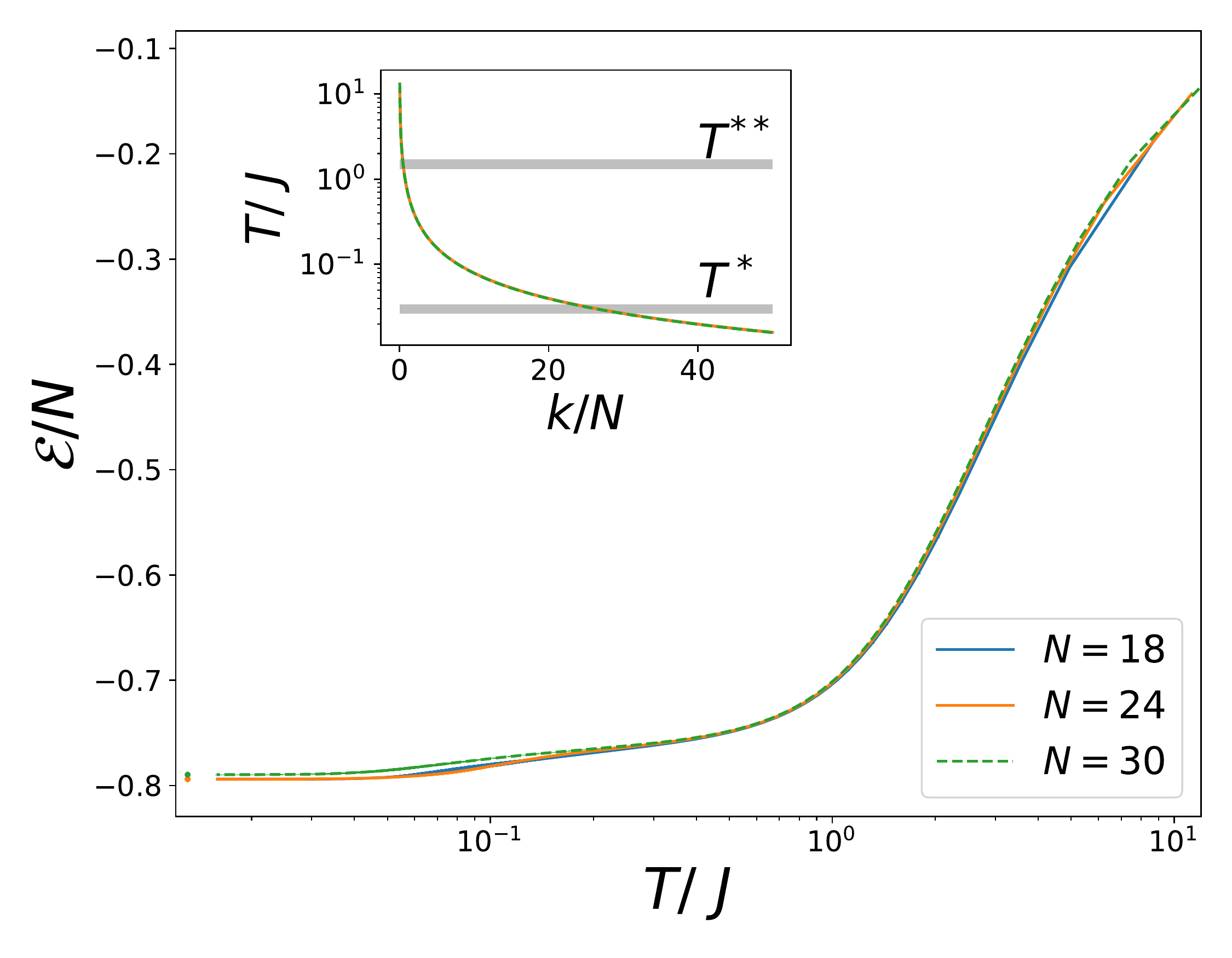}
  \caption{Internal energy in the Kitaev model with $N=18, 24$ and $30$.
    The inset shows the temperature as a function of $k$.
    The results are obtained from the TPQ state generated from $100$ and $25$ samples of the initial random state 
    for $N=18, 27$, and $N=30$, respectively, 
    and the error bars stand for the standard deviation of the results.
    Circles represent the ground state energies for the corresponding system.
  }
  \label{Kitaev_energy}
\end{figure}
In fact, we find in Fig.~\ref{Kitaev_energy}
two shoulder structures appear in the internal energy
around $T\sim 0.1J$ and $\sim 0.8J$.
This behavior is clearly found as a double-peak structure in the specific heat~\cite{Nasu},
and these peaks are located at $T^*/J\sim 0.03$ and $T^{**}/J\sim 1.5$.
The characteristic temperature $T^*$ is relatively low due to this fractionalization phenomenon.
The inset of Fig.~\ref{Kitaev_energy} shows that 
the temperature as a function of $k/N$.
We find that the curve of the temperatures is well scaled by $k/N$,
which is similar to that for the KH model.
Therefore, the TPQ state at the lower characteristic temperature $T=T^*$ 
is obtained with $k\sim 30N $
when appropriate parameters are given.

\bibliography{./refs}

\end{document}